\documentstyle[12pt,aaspp4]{article} 
\newcommand{\postscript}[2] 
{\setlength{\epsfxsize}{#2\hsize} 
\centerline{\epsfbox{#1}}} 
\def\tempest%
{\begin{array}{ccc} 
1 & 1 & 1 \\ 
1 & 1 & 1 \\ 
4 & 3 & 8 
\end{array}}

\def\e{{\rm E}}
\def\btheta{\hbox{$\theta\hskip-6.3pt\theta$}}

\begin{document}

\title{Astrometric Resolution of Severely Degenerate Binary
Microlensing Events} 
\author 
{Andrew Gould}
\affil{Ohio State University, Department of Astronomy, 
140 W.\ 18th Ave., Columbus, OH 43210, USA; gould@astronomy.ohio-state.edu} 
\author 
{Cheongho Han}
\affil{Department of Astronomy \& Space Science, Chungbuk National University,
Chongju, Korea 361-763; cheongho@astronomy.chungbuk.ac.kr}

\begin{abstract} 

	We investigate whether the ``close/wide'' class of degeneracies in 
caustic-crossing binary microlensing events can be broken astrometrically.  
Dominik showed that these degeneracies are particularly severe because they 
arise from a degeneracy in the lens equation itself rather than a mere 
``accidental'' mimicking of one light curve by another.  A massive observing 
campaign of five microlensing collaborations was unable to break this 
degeneracy photometrically in the case of the binary lensing event MACHO 
98-SMC-1.  We show that this degeneracy indeed causes the image centroids
of the wide and close solutions to follow an extremely similar pattern of
motion during the time when the source is in or near the caustic.  
Nevertheless, the two image centroids are displaced from one another and
this displacement is detectable by observing the event at late times.
Photometric degeneracies therefore can be resolved astrometrically, even
for these most severe cases.


\end{abstract} 

\section{Introduction} 

	Caustic-crossing binary microlensing events are potentially very 
useful, but their interpretation can be problematic.  If there is good
photometric coverage of a caustic crossing, one can measure the limb
darkening of the source (Albrow et al.\ 1999b; Afonso et al.\ 2000;
Albrow et al.\ 2000), and if this is combined with spectroscopy, one can
resolve the source's spectral features as a function of angular position
(Gaudi \& Gould 1999).  If there is sufficiently good coverage of the
event to obtain a {\it unique} binary solution, then one determines
the binary mass ratio and in some cases other information about the binary
(Albrow et al.\ 2000).  If this information can be obtained for a number
of events, then one can infer statistical properties about the binaries
in the lens population as a whole (Gaudi \& Sackett 2000).

	In some cases, it is possible to measure the proper motion $\mu$ 
of the lens relative to the observer-source line of sight.  This requires 
three pieces of information from the photometric light curve.  First, one must
measure the time it takes the source star to cross the caustic, $\Delta t$ 
which can be done from photometry of the caustic-crossing alone (e.g.\ 
Albrow et al.\ 1999a,c).  Second, one must measure the angle $\phi$ of this 
crossing, which requires a {\it unique} binary solution for the event as 
a whole.  Third, one must determine the angular size $\theta_*$ of the 
source from its color and apparent magnitude using an empirically calibrated 
relation (van Belle 1999).  The color is quite easily measured but the 
apparent magnitude again requires a {\it unique} binary solution.  The proper 
motion is then $\mu=\theta_*/(\Delta t|\sin\phi|)$.  Five groups (Afonso 
et al.\ 1998; Udalski et al.\ 1998; Alcock et al.\ 1999; Albrow et al.\ 
1999; Rhie et al.\ 1999) combined observations from 8 observatories to 
measure the proper motion of MACHO 98-SMC-1 and so proved beyond reasonable 
doubt that the lens was in the SMC and not in the Galactic halo (Afonso 
et al.\ 2000).

	As can be seen from this brief summary, many applications of binary
lenses require that one obtain a unique binary-lens solution to the event.
However, Dominik (1999a) presented multiple solutions to a number previously 
published events and argued that degeneracies of this sort may be generic. 

	Han, Chun, \& Chang (1999) therefore investigated whether such 
degeneracies can be broken astrometrically.  In a binary lensing event 
there are three or five images depending on whether the source is outside 
or inside the caustic.  The combined light of these three or five images
makes up the photometric light curve which is the only effect that has been
observed to date.  The images are separated by of order the Einstein radius,
$\theta_\e$ which is a few hundred $\mu$as for typical events.  Hence, the 
images cannot be separately resolved with any existing or planned instrument.  
However, the image centroid deviates from the source position by a vector 
amount $\delta\btheta_c=(\delta\theta_{c,x},\delta\theta_{c,y})$ which is 
also of order $\theta_\e$.  The Space Interferometry Mission (SIM) with its
planned $\sim 4\,\mu$as precision will therefore have the capability to measure
this deviation, and several ground-based interferometers may also achieve
the necessary precision.

	Han et al.\ (1999) explicitly showed that the four solutions
that Dominik (1999a) presented for OGLE-7 (Udalski et al.\ 1994), which
all fit the observed light curve extremely well, had radically different
astrometric trajectories.  Hence, had there been astrometric data, this
degeneracy could have easily been broken.

	Subsequently, Albrow et al.\ (1999c) developed a general method
for finding solutions in events with well-covered caustic crossings, and
Afonso et al.\ (2000) applied this method to MACHO 98-SMC-1 and found two 
solutions that fit the full data set equally well.  See Figures 1 and 2, 
below. In spite of Dominik's (1999a) work showing that degeneracies in 
earlier light curves were common, the degeneracy in MACHO 98-SMC-1 came 
as something of a surprise because the 5-collaboration data set was far 
superior to those of the events investigated by Dominik (1999a).

	However, simultaneously with Afonso et al.'s (2000) {\it empirical} 
discovery of a severe degeneracy in MACHO 98-SMC-1, Dominik (1999b) found 
{\it an entire class} of severe degeneracies between ``close'' and ``wide'' 
binaries, i.e., binaries with projected angular separations small and large 
compared to $\theta_\e$.  Indeed, MACHO 98-SMC-1 turns out to be a particular 
case of this class.

	The disturbing thing about the Dominik (1999b) close/wide 
degeneracies, and what makes them so difficult to break, is that they 
derive from a degeneracy in the lens equation itself.  That is, while some 
of the degeneracies found by Dominik (1999a) may be regarded as due 
``accidental'' similarities between different light curves (the sum of 
the magnification of the three or five images), the close/wide degeneracies 
are rooted in the similarities of the individual images.  This immediately 
raises the question of whether it is possible to break these degeneracies 
at all, even using astrometric data as Han et al.\ (1999) showed could be
done for the earlier (Dominik 1999a) degeneracies.  We address that question
here.

\section{Astrometric Resolution}

	To investigate this question, we examine the astrometric behavior
of the two solutions\footnote{Afonso et al.\ (2000) actually found two 
wide solutions, a static and a rotating one.  The static solution was 
completely consistent with the data in the neighborhood of the observable 
event, but was by chance ruled out by early data about 500 days before the 
event.  For simplicity, and because we are trying to illustrate a general 
principle rather than specifically investigate MACHO 98-SMC-1, we will use 
the static wide solution and therefore will ignore the early data.  This 
will allow us to compare two static systems, thus ensuring that differences 
between the astrometric trajectories are not due to the fact that one is 
rotating and the other is not.} to MACHO 98-SMC-1 found by Afonso et al.\ 
(2000).  The Einstein radii $\theta_\e$ are $74\,\mu$as and $167\,\mu$as, 
the Einstein crossing times $t_\e$ are 99 days and 165 days, the mass
ratios $M_2/M_1$ are 0.50 and 4.17, and the separations are $d\theta_\e$ 
where $d=0.54$ and $3.25$.  Here $M_1$ is the mass of the component that is 
closer to the caustic that the source passes through.  The full solutions 
are described in Tables 1 and 2 of Afonso et al.\ (2000).

	Figure 1 shows the trajectory of the source relative to the binary 
in the two solutions.  Figure 2 is adapted from Figures 3 and 4 of Afonso 
et al.\ (2000) and shows the predicted light curves in $I$ band for these 
two solutions.  Time is shown as HJD$'$=HJD-2450000.  The data are binned 
in 1-day intervals except in the immediate neighborhood of the caustics 
where there are 0.1-day bins.  Data taken in other bands are adjusted to 
the $I$ band system using the source and background fluxes from each 
observatory and each band as determined from the overall fit.  See Afonso et 
al.\ (2000).  The main conclusion from Figure 2 is that the two solutions 
are essentially identical from a photometric standpoint.

	Figure 3 shows the astrometric deviation of the light centroid from 
the source position for the close and wide solutions, respectively.
The dashed  portions of the curve show the jumps at the 
times of the caustic crossings.  These jumps would be discontinuous for a 
point source, but in fact take place by a (rapid) continuous motion for a 
finite source.  Note that the two displacement curves look extremely similar.  
This similarity derives from the underlying degeneracy in the lens equation 
that was discovered by Dominik (1999b).

	Although the {\it pattern} of centroid motion is extremely similar 
in the two cases, the two curves are actually displaced from one another by 
an offset
\begin{equation}
\Delta\delta\btheta_c(t) = \delta\btheta_{c,\rm close}(t)-
\delta\btheta_{c,\rm wide}(t),
\label{eqn:deldel}
\end{equation}
which is about $40\,\mu$as, i.e., $\sim 0.5\,\theta_\e$ for the close binary 
or by $\sim 0.25\,\theta_\e$ for the wide binary.  Such an offset is not 
observable if the observations are restricted to times when $\Delta\delta
\btheta_c$ is approximately constant.  However, at very early or very late 
times, $\delta\btheta_c\rightarrow 0$, since when the source is far from the 
lens, the image and source positions coincide.  Because this occurs for both 
models, $\Delta\delta\btheta_c$ must vanish at these times.  How long, in 
practice, must one wait to tell the difference between the two models?

	This question is addressed in Figure 4 where we plot the offset 
between the two models, $\Delta\delta\btheta_c(t)$ as a function of time.  
The main figure shows the behavior of $\Delta\delta\btheta_c(t)$ over the 
whole event, while the
inset is restricted to times during and after the caustic 
crossing (when in practice astrometric observations might first be triggered).
The offset shows some structure on scales of $\sim 5\,\mu$as during the time 
when the source is inside the caustic, but the main thing to notice is that 
over the next year it changes by $20\,\mu$as and thus would be noticed 
if the event were monitored with SIM-like precision. The full $40\,\mu$as 
change would take place only after about a decade.  Note that to the extent 
that $d\Delta\delta\btheta_c/d t$ can be approximated as a constant, 
$\Delta\delta\btheta_c$
 cannot be detected at all, because such uniform motion can be subsumed in 
the fit for the proper motion of the source.  However, from Figure 4, 
$\Delta\delta\btheta_c(t)$ slows down dramatically after about 1 year, so 
that after 2 years, its non-uniform motion could be unambiguously 
distinguished from uniform source motion.

	It is also instructive to look at the behavior of Figure 4 at early 
times.  Of course no astrometric measurements could have been taken then 
because there had been no signature of an event.  However, the entire event 
could have just as well taken place in reverse.  In this case {\it 
post-caustic} astrometric measurements would probe dramatic changes in the 
offset, much larger than the $40\,\mu$as changes in the event proceeding in 
its actual direction.  The reason for this can be seen in Figure 1: 
the source passes relatively close to the companion binary 
member and this passage induces a large astrometric deviation.  (Indeed, this
passage is so close that it induces a noticeable deviation in the photometric
light curve which is why the early data for this event ruled out the static 
wide solution.)\ \ In general, the source is not likely to pass close to both 
members, so that deviations of the type seen in Figure 4 are unlikely.  

	However, the $\sim 40\,\mu$as offset seen in Figure 3 at times when 
the source is inside the caustic is a generic feature of this caustic and 
does not depend in any way on the direction of the source trajectory through 
the caustic.  Therefore, it is generically possible to break the close/wide 
(Dominik 1999b) degeneracy astrometrically, even when it is extremely 
difficult to do so photometrically.

\section{Discussion}

	As a practical matter, SIM could not resolve the degeneracy in
MACHO 98-SMC-1 because the source is only $I\sim 22$, far too faint for SIM 
to follow.  Most events that SIM could monitor would be in the bulge where
there are far more events and where the sources are much brighter.  For
these events, the typical Einstein radius is probably 
$\theta_\e\sim 300\,\mu$as, so the astrometric deviations would be several
times larger than for MACHO 98-SMC-1.  Hence, it seems likely that for
sources that could be monitored astrometrically at all, breaking the
degeneracy would be well within SIM's capabilities.

	Finally, we ask: what is the fundamental physical reason that
the photometric degeneracy is reproduced as an astrometric degeneracy in the
neighborhood of the caustic, but can be broken astrometrically at late times?
This can most easily be seen by looking at Figure 1.  In each
model, the caustic that is crossed is associated with the mass at the right.
The Einstein crossing times associated with these masses, 
$t_\e'=[M_1/(M_1+M_2)]^{1/2}t_\e$,
are about the same in the two cases, $t'_{\e,\rm close}=81\,$days and  
$t'_{\e,\rm wide}=72\,$days,
respectively.  We therefore show the size of mass $M_1$ to be
$t'_{\e,\rm close}/t'_{\e,\rm wide}=1.25$ times larger in the close-binary
panel than in the wide-binary panel.
The lens equation is very similar in the neighborhood of this
mass, which is the origin of both the photometric and astrometric degeneracy.
However, for the wide-binary solution, the very large mass of the companion
at the left displaces the entire image structure to the right by
$\sim [M_2/(M_1+M_2)]^{1/2}\theta_\e/d\sim 57\,\mu$as.  This displacement
only gradually returns to zero: even at time $d t_\e \sim 540\,$days after
the event, it has only fallen by half.  By contrast, once the source has
left the vicinity of $M_1$ of the close binary, there are no large and 
distant masses that could significantly displace the images relative to the
source.

\smallskip

{\bf Acknowledgements}: We thank Scott Gaudi for stimulating discussions.
Work by AG was supported in part by grant AST 97-27520 from the NSF.
Work by CH was supported by grant KRF-99-041-D00442 of the Korea Research
Foundation.  CH thanks the Ohio State University Astronomy Department
for its hospitality during a visit during which most of the work on this 
paper was completed.

\bigskip

\clearpage

\postscript{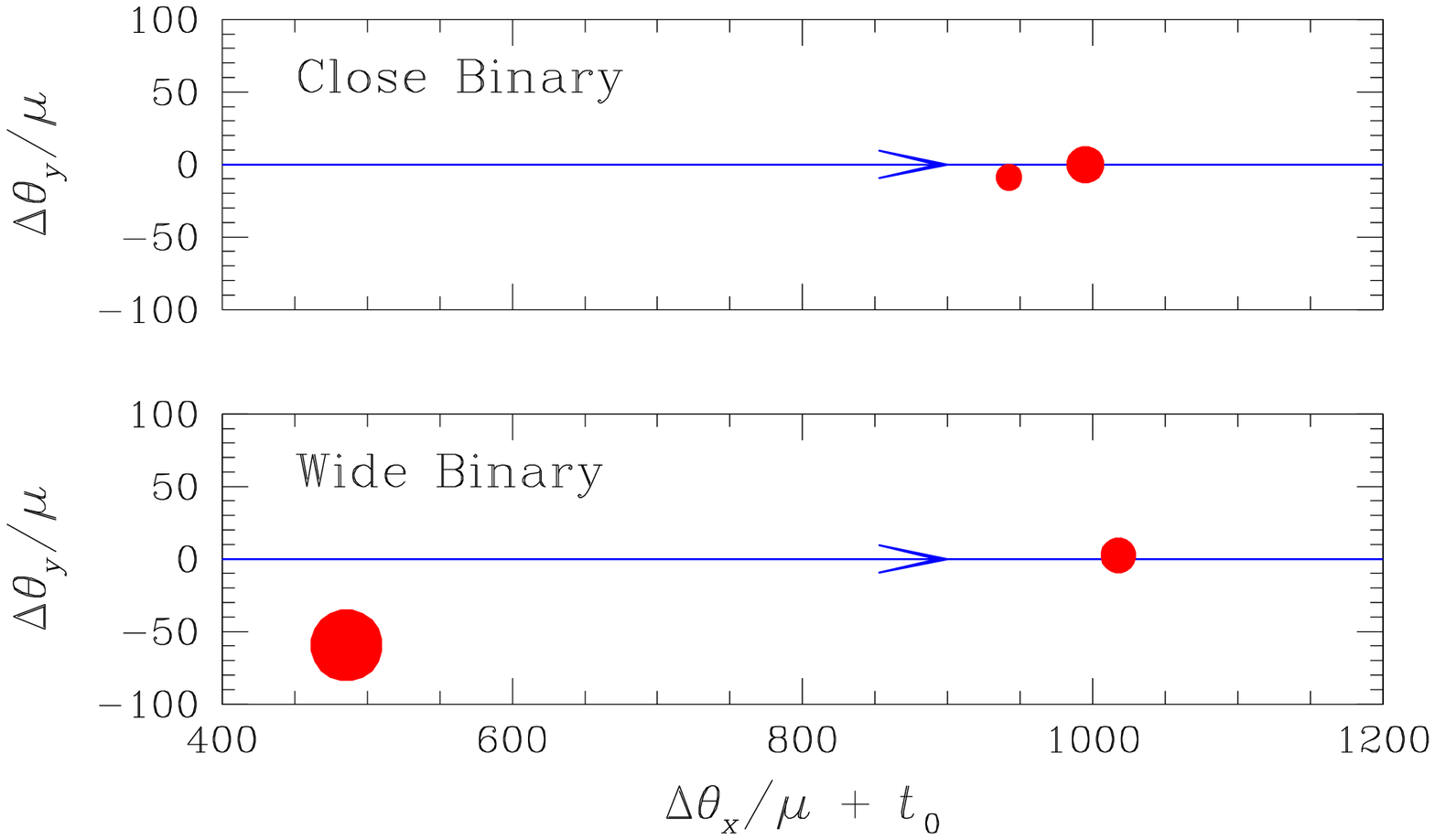}{1.0}
\vskip-1.5cm
\noindent
{{\bf Figure 1:}\ 
Positions of the components of the binary lens MACHO 98-SMC-1 in the two 
models of Afonso et al.\ (2000).  The size (area) of the dots indicates 
the relative masses of the components.  The panels show angular position
scaled by the proper motion $\mu$.  The units are therefore time which
means that the source trajectory is shown as a function of 
$\Delta\theta_x/\mu+t_0 =$ HJD$'$, and is therefore the same in the two panels.
}

\postscript{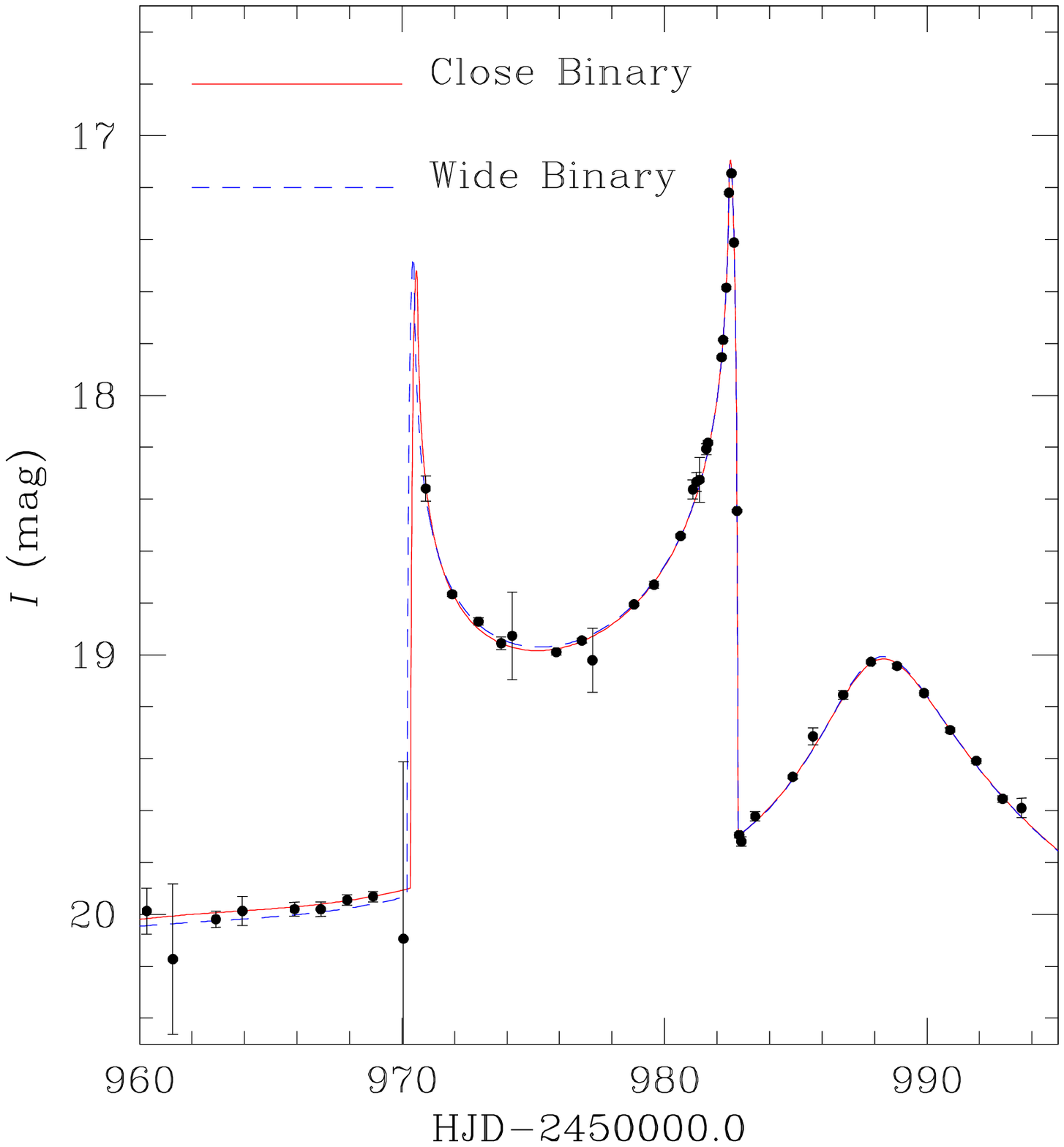}{1.05}
\noindent
{{\bf Figure 2:}\ 
Light curves for close binary ({\it solid}) and wide binary ({\it dashed})
models for the caustic-crossing binary microlensing event MACHO 98-SMC-1,
together with binned data from 5 microlensing collaborations.  Both the
curves and data are adapted from Figures 3 and 4 of Afonso et al.\ (2000).  
The non-$I$ points have been put on the $I$ band system using the solutions
of Afonso et al.\ (2000) so that both curves could be shown on the same plot.
The two models predict virtually identical photometric results.
}\clearpage

\postscript{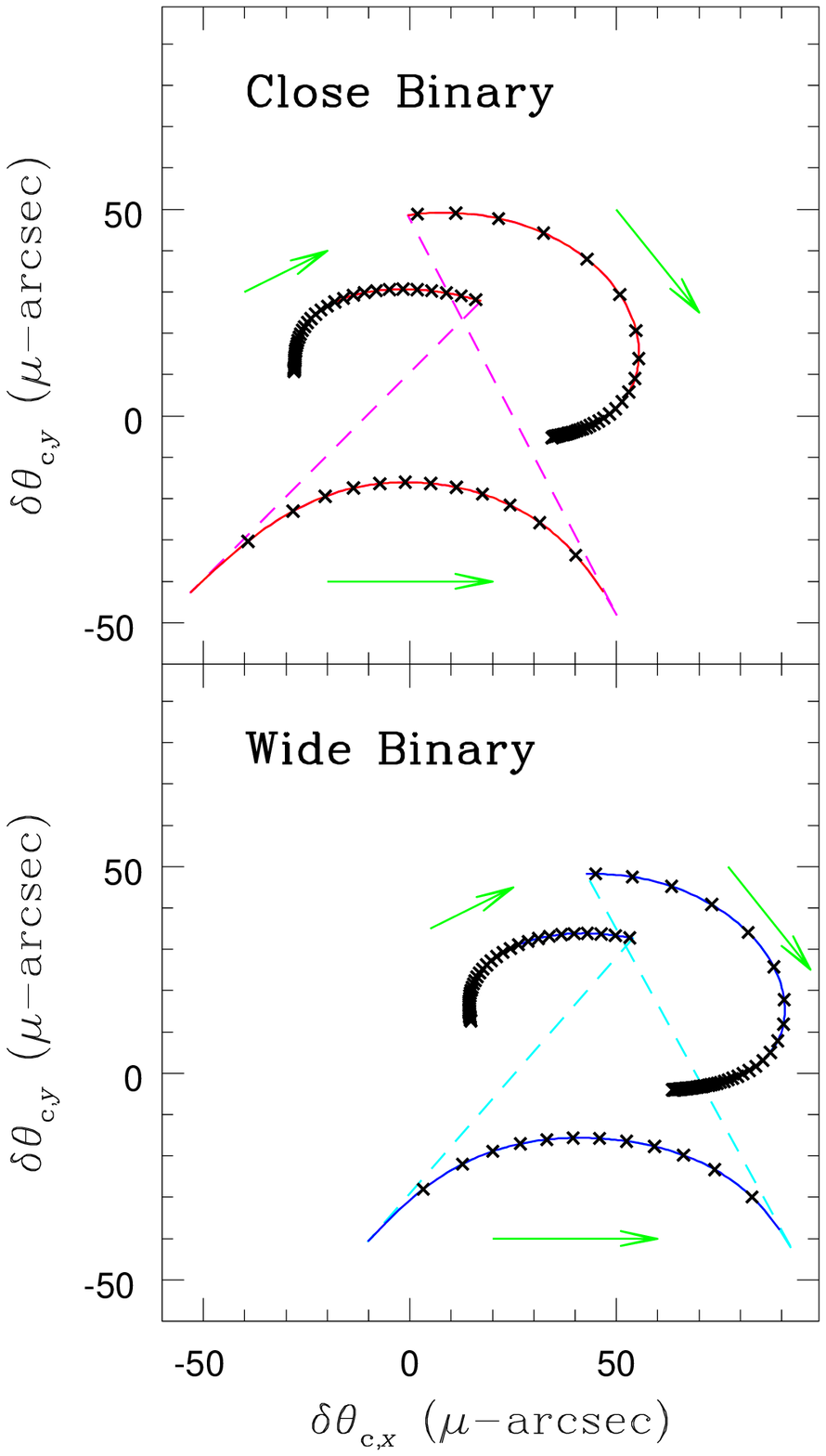}{0.85}
\noindent
{{\bf Figure 3:}\ 
Astrometric deviation $\delta\btheta_c$ of the image centroid from the source
position for the same two models shown in Figure 2 and over the
same time interval, $960\leq{\rm HJD}'\leq 995$.  The crosses show the progress
of the event in 1 day intervals, and the arrows designate the direction of the 
centroid motion.  The dashed lines show the ``instantaneous jumps'' that the 
image centroid of a point source would undergo at the caustic crossing.  Finite 
source effects (not shown) would make these transitions continuous and would 
fore-shorten them by about 3\%.  The arcs at the bottoms represent the image 
centroid positions when the source is inside the caustic.  The pattern of 
motion in the two cases looks extremely similar, confirming the photometric 
degeneracy illustrated in Figure 2.  However, the two trajectories are offset 
by $\sim 40\,\mu$as, meaning that they can be distinguished if the zero-point
of astrometry is established by sufficiently late-time observations.
}\clearpage

\postscript{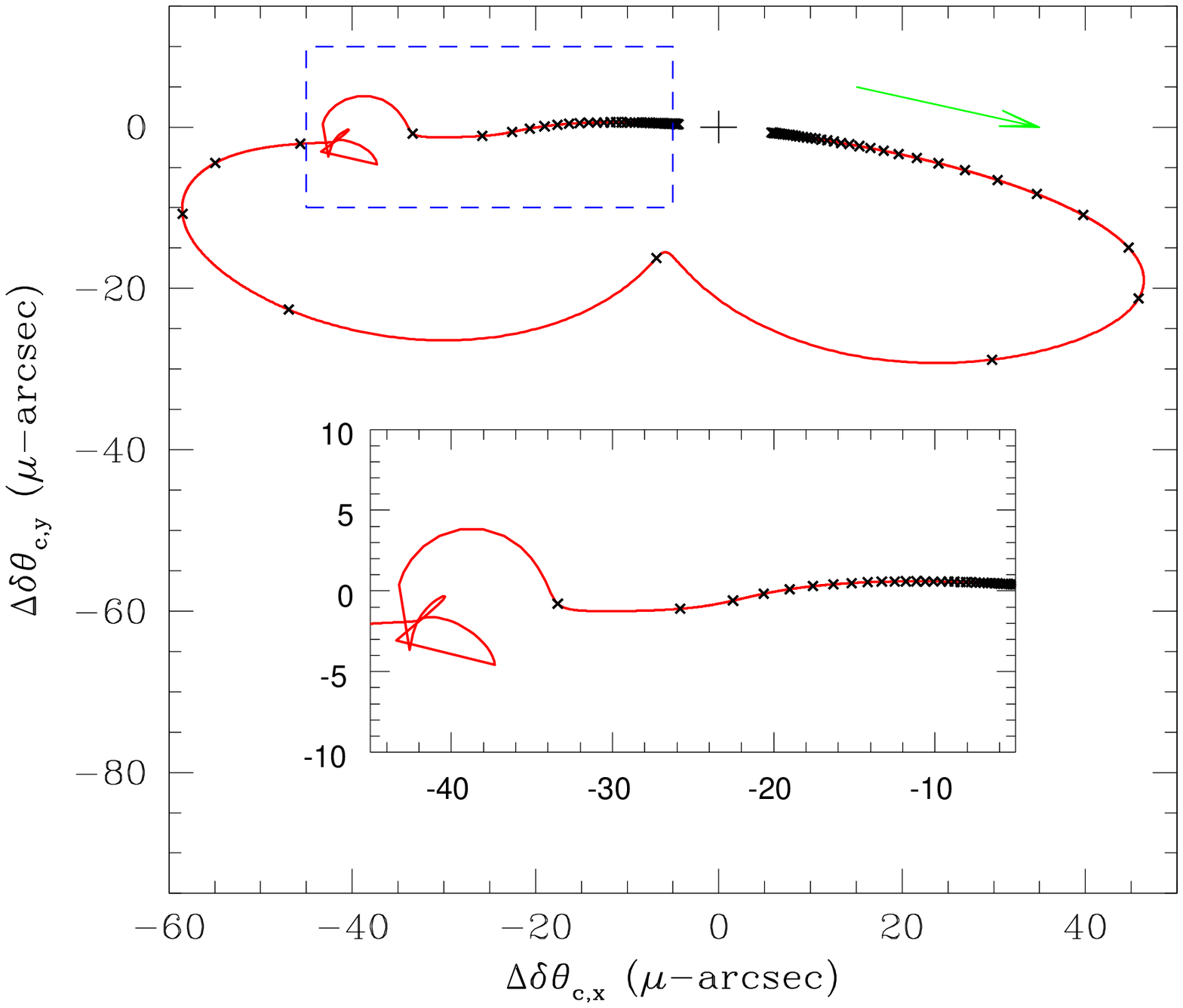}{0.9}
\vskip-0.7cm
\noindent
{{\bf Figure 4:}\ 
Difference $\Delta\delta\btheta_c$ between the two astrometric deviations
shown in Figure 3, with evaluations every 100 days shown by
crosses.  The full figure shows $\Delta\delta\btheta_c$ over a period of 
20 years, while the inset shows only the period during and after the
caustic crossing (when astrometric measurements might reasonably have been
triggered).  During the year after the caustic crossing, 
$\Delta\delta\btheta_c$ changes by $\sim 20\,\mu$as.  To the extent that
this is consistent with uniform motion, it could not be disentangled from
the uniform proper motion of the source.  However, during the subsequent
year, $\Delta\delta\btheta_c$ slows down substantially, so that its motion
over 2 years could easily be distinguished from uniform motion.  Thus,
astrometric measurements would distinguish between the two solutions.
}


\begin{references} 
\reference{} Afonso, C.\ et al.\ 1998, \aap, 337, L17 
\reference{} Afonso, C.\ et al.\ 2000, \apj, 532, 000 (astro-ph/9907247)
\reference{} Albrow, M.\ et al.\ 1999a, \apj, 512, 672 
\reference{} Albrow, M.\ et al.\ 1999b, \apj, 522, 1011 
\reference{} Albrow, M.\ et al.\ 1999c, \apj, 522, 1022 
\reference{} Albrow, M.\ et al.\ 2000, \apj, 534, 000 (astro-ph 9910307)
\reference{} Alcock, C.\ et al.\ 1999, \apj, 518, 44
\reference{} Dominik, M.\ 1999a, \aap, 341, 943
\reference{} Dominik, M.\ 1999b, \aap, 349, 108 
\reference{} Gaudi, B.S., \& Gould, A.\ 1999, \apj, 513, 619
\reference{} Gaudi, B.S., \& Sackett, P.D. 2000, \apj, 529, 000
(astro-ph/9904339)
\reference{} Han, C., Chun, M.-S., \& Chang, K.\ 1999, \apj, 526, 405
\reference{} Rhie, S.H., Becker, A.C., Bennett, D.P., Fragile, P.C., 
Johnson, B.R., King, L.J., Peterson, B.A., \& Quinn., J.\ 1999, 
\apj, 522, 1037
\reference{} Udalski, A., Szyma\'nski, M.,  Mao.\ S., Di Stefano, R.,
Kaluzny, J., Kubiak, M., W., Mateo, M., \& Krzemi\'nski, M.\ 
1994, \apj, 436, L103 
\reference{} Udalski, A., Szyma\'nski, M., Pietrzy\'nski, G., Kubiak, M., 
Wo\'zniak, 
P., \& \.Zebru\'n, K.\ 1998, AcA, 48, 431
\reference{} van Belle, G.T.\ 1999, \pasp, 111, 1515
\end{references}
\end{document}